\documentclass[prl,twocolumn,superscriptaddress]{revtex4}
\usepackage{}%
\usepackage{amsfonts}
\usepackage{amsmath}
\usepackage{amssymb}
\usepackage{graphicx}%
\usepackage{color}
\emergencystretch=\maxdimen
\begin{document}

\title{Lattice thermal conductivity of ZrSe$_2$ based on the anharmonic phonon approach and on-the-fly machine learning force fields}

\author{Yong Lu}
\thanks{Corresponding author; luy@mail.buct.edu.cn}
\affiliation{College of Mathematics and Physics, Beijing University of Chemical Technology, Beijing 100029, China}
\author{Fawei Zheng}
\thanks{Corresponding author: fwzheng@bit.edu.cn}
\affiliation{Centre for Quantum Physics, Key Laboratory of Advanced Optoelectronic Quantum Architecture and Measurement(MOE),
School of Physics, Beijing Institute of Technology, Beijing, 100081, China}

\date{\today}
\clearpage

\begin{abstract}
The lattice thermal conductivity (LTC) of ZrSe$_2$, a typical layered transition metal disulfide, has been calculated using a hybrid approach that combines force field molecular dynamics (MD) simulation and Boltzmann transport equation (BTE). In this approach, the phonon quasiparticle picture of each normal mode can be obtained directly by the velocity autocorrelation function and its power spectrum projected in $q$-space. By employing the retarded one phonon Green's function method, the phonon quasiparticle frequency and lifetime of each independent normal mode are effectively determined. On-the-fly machine learning force fields combine the precision of quantum mechanics and the scale of classical MD to analyze sizable supercells with long-wavelength phonons. This yields accurate LTC by using sufficient $q$-samples in the Brillouin zone, which cannot be achieved at the \emph{ab initio} molecular dynamics scale. The convergent LTC tensor of bulk and monolayer ZrSe$_2$ decays faster with increasing temperature than the well-known $\frac{1}{T}$ scale, which is typically observed when considering only three-phonon scattering. The phonon lifetime and mean free path exhibit significant dependence on temperature. MD simulations encompass all orders of anharmonic effects, thereby enabling an accurate description of anharmonic interactions between phonons at finite temperatures. Moreover, this approach respects the contribution of each normal mode to the LTC based on the BTE, which facilitates the quantitative analysis of phonon anharmonic properties and the role of specific normal modes.
\end{abstract}

\maketitle
\clearpage
\section{I. INTRODUCTION}
Thermal transport is a fundamental property of condensed matter, in which heat is transmitted through microscopic collisions of particles such as phonons and electrons \cite{book1}. In nonmetallic solids, phonons are the primary heat carriers, and anharmonic inelastic collisions between them significantly impact thermal conductivity at temperatures close to or above room temperature. The study of phonon anharmonic behavior has always been a challenge in condensed matter theory. Over time, numerous spectral phonon analysis methods have emerged and been employed to investigate phonon scattering \cite{Klemens1,Klemens2,Klemens3,Herring,Casimir,Berman1,Berman2,Callaway,Allen}. However, these theoretical models often incorporate empirical fitting parameters and lack predictive power, thus hindering their widespread applicability.

The development of numerical methods based on the Boltzmann transport equation (BTE) and molecular dynamics (MD) simulation has improved the scope of application by predicting lattice thermal conductivity (LTC) from the atomic structure without relying on empirical fitting parameters. Generally, the phonon BTE \cite{PBTE1,PBTE2,PBTE3} is based on the first-principles approach and requires only the harmonic and anharmonic interatomic force constant matrix as input. However, the cost of calculation is high, and typically only third-order interactions are used to handle anharmonic phonon interactions. While third-order interactions performed at the ground state can capture essential phonon anharmonicity features, they may not fully describe phonon quasiparticles at high temperatures where higher-order processes can have a significant impact on heat transfer \cite{Feng2017,Xie2020}. Furthermore, this method is limited in dealing with large or complex systems due to the scale of simulations.

The MD method presents an alternative solution for the calculation of heat transfer, as it considers all orders of anharmonic terms and scales better with the system size. Equilibrium MD based on the Green-Kubo method \cite{Green1,Green2,Kubo1,Kubo2,EMD1,EMD2} and non-equilibrium MD \cite{NEMD1,NEMD2,NEMD3} based on Fourier's law are two accessible ways to evaluate LTC. However, its limitation lies in the irreconcilability of computational accuracy and simulation scale for a long time. In classical MD, the calculation of particle velocity is dependent on the calculation of position and potential energy, leading to numerical errors due to the insufficient accuracy of interatomic potentials. Although \emph{ab initio} molecular dynamics (AIMD) simulations can provide potential with quantum mechanical accuracy, they are constrained by simulation scale. Fortunately, the regression technique based on machine learning (ML) has emerged as a promising tool for building interatomic potentials, which can combine the accuracy of quantum mechanics with the computational scale and efficiency of classical potentials \cite{Schmidt2019,Hu2020,Chakraborty2020,Chowdhury2020,Fan2021}. The success of ML technology is contingent upon the availability of data that can be used to train and model the force field of a particular crystal structure. Typically, these training datasets are derived from AIMD, resulting in the force field potential model inheriting the precision of these calculations. To reduce human intervention, the on-the-fly machine learning force field (MLFF) has gained more attention \cite{Zhenwei2015,Jinnouchi2019,Gilad2020}. It enables real-time calculation of each step of the MD simulation, eliminating the need for pre-trained force field model and accelerating the calculation process.

Nevertheless, the MD method is based on thermodynamic statistical theory, and it cannot evaluate the contribution of each normal mode to heat transport, limiting quantitative analysis of phonon anharmonicity and the action of specific normal modes. Due to the advantages of accuracy and scale offered by MLFF, it brings more possibilities for us to develop new methods for calculating LTC. In this paper, we use a hybrid method that effectively combines the large-scale force field molecular dynamics (FFMD) and the BTE to calculate the LTC. The existence of phonon quasiparticles is a prerequisite for performing thermodynamic property calculations. In this method, phonon quasiparticles can be directly observed and defined using the velocity autocorrelation function (VAF) and power spectrum based on the FFMD simulation. The relationship between the VAF and the phonon quasiparticle frequency and lifetime is established based on the retarded phonon Green's function. By projecting the real-space VAF onto all individual modes, temperature-dependent phonon frequencies and lifetimes for all normal modes can be obtained, which are further used to calculate the LTC based on the BTE. It is worth noting that the mode-projected VAF depicts how one mode interacts with all the other modes that are sampled in the MD simulation. Hence, the number of $q$ samples is directly related to the number of primitive cells in the supercell used in the FFMD simulation. By implementing FFMD, we are capable of analyzing sizable supercells with long-wavelength phonons, which guarantees adequate $q$-sampling in the Brillouin zone (BZ) and engenders precise LTC computations that cannot be achieved through AIMD. We chose the transition metal disulfide ZrSe$_2$ as the research carrier. ZrSe$_2$ has a moderate bandgap of about 1.0--1.2 eV \cite{Moustafa2009,Moustafa2016}, which makes it suitable for optoelectronic semiconductor devices due to its excellent light absorption properties in the visible to near-infrared spectral range. It also exhibits excellent electron transfer rates, making it suitable as an electron transport layer material in micro and nanoelectronic devices, as well as a thermoelectric resistor material in thermoelectric generators. Therefore, an effective prediction of its LTC at finite temperatures is of great significance in controlling the heat dissipation and maintaining the performance stability of high-precision microelectronic devices. We calculated the LTC of bulk phase and monolayer ZrSe$_2$ at finite temperatures. The temperature-dependent anharmonic phonon properties, including the frequency, the phonon lifetime, and the phonon mean free path (MFP), were also predicted and discussed. This approach surpasses the limitations of previous BTE methods, and enhances the precision of classical MD, presenting a promising universal method for calculating the LTC of crystals.

\section{II. Methods and Computational Details}
\subsection{Anharmonic phonon approach}
For a lattice system, the retarded one phonon Green$'$s function of normal mode ($\textbf{q}, s$), denoted as $\lambda$ for short, can be defined as \cite{Sun2010}
\begin{equation}
G_{R}(\lambda,t)=\frac{1}{i\hbar}\langle [u_{\lambda}(t),u_{\lambda}^{\dagger}(0)]\rangle \theta(t),
\end{equation}
where $u_{\lambda}=\sqrt{\hbar/(2\omega_{\lambda})}(a_{\lambda}+a_{-\lambda}^{\dagger})$ denotes the normal mode coordinates, $\hbar$ is the reduced Plank constant, $\omega_{\lambda}$ is the phonon angular frequency, $a_{\lambda}$ and $a_{-\lambda}^{\dagger}$ are the annihilation and generation operators, and $\theta(t)$ is the step function, respectively. Its Fourier transform $G_{R}(\lambda,\omega)$ can be represented in terms of phonon self-energy $\Sigma(\lambda,\omega)$ as
\begin{equation}
G_{R}(\lambda,\omega)=\frac{1}{\omega^{2}-\omega_{\lambda}^{2}-2\omega_{\lambda}\Sigma(\lambda,\omega)},
\end{equation}
where $\Sigma(\lambda,\omega)=\Sigma_{1}(\lambda,\omega)+i\Sigma_{2}(\lambda,\omega)$. Assuming that the frequency dependence of $\Sigma(\lambda,\omega)$ is weak and $|\Sigma(\lambda,\omega)| \ll \omega_{\lambda} $, the real part $\Sigma_{1}(\lambda,\omega)=\Delta \omega_{\lambda}$ corresponds to the frequency shift with respect to the harmonic frequency $\omega_{\lambda}$ and the imaginary part $-\Sigma_{2}(\lambda,\omega)=\Gamma_{\lambda}$ corresponds to the phonon linewidth \cite{Ladd,Sun2010,Zhang2014}. In this case, Eq. (2) can be simplified as
\begin{equation}
G_{R}(\lambda,\omega)\approx -\frac{1}{\widetilde{\omega}_{\lambda}}\left[\frac{\widetilde{\omega}_{\lambda}}{\widetilde{\omega}_{\lambda}^{2}-(\omega+i\Gamma_{\lambda})^{2}}\right],
\end{equation}
where $\widetilde{\omega}_{\lambda}=\omega_{\lambda}+\Delta \omega_{\lambda}$ is the phonon quasi-particle frequency.

In the classical limit, the one phonon Green$'$s function can be connected with the classical correlation function in the frequency domain by the Kubo transform \cite{Kubo1},
\begin{equation}
G_{C}(\lambda,\omega)=-\beta\langle v_{\lambda}^{*} u_{\lambda} \rangle_{\omega} = \frac{\beta}{i\omega}\langle v_{\lambda}^{*} v_{\lambda} \rangle_{\omega}.
\end{equation}
It is noted that the expression in the bracket of Eq. (3) corresponds to the Fourier transform of the function $\textup{sin}(\widetilde{\omega}_{\lambda}t)\textup{exp}(-\Gamma_{\lambda}t)\theta(t)$. By comparing Eqs. (3) and (4), we can extract the expression of the VAF:
\begin{equation}
\langle v^{*}_{\lambda}(0) v_{\lambda}(t)\rangle=k_{B}T\textup{cos}(\widetilde{\omega}_{\lambda}t)\textup{exp}(-\Gamma_{\lambda}t),
\end{equation}
which shows the damped harmonic oscillator with an associated exponential function describing the mode lifetime. Generally, the VAF for crystalline systems is defined as \cite{Wang1990},
\begin{equation}
\langle v^{*}_{\lambda}(0) v_{\lambda}(t)\rangle=\lim\limits_{t_{0}\rightarrow{\infty}} \frac{1}{t_0} \int^{t_{0}}_{0} v_{\lambda}(t')v_{\lambda}(t'+t)dt'.
\end{equation}
And the mode-projected velocity, $v_{\lambda}(t)$, can be expressed as
\begin{equation}
v_{\lambda}(t)=\sum_{j=1}^{N}\sqrt{M_j} v_{j}(t)\textup{exp}(-i\textbf{q}\cdot \textbf{R}_{j})\cdot \hat{{e}}_{\lambda},
\end{equation}
where $v_{j}(t)$ ($j=1,...,N$) are atomic velocities obtained from MD with $N$ atoms per supercell. $\hat{{e}}_{\lambda}$ is the polarization vector of the harmonic phonon mode $\lambda$, which can be routinely obtained by performing phonon calculations with standard approaches such as the finite displacement method and the density functional perturbation theory~\cite{Baroni1987,Baroni2001,Togo2010}. $M_j$ and $\textbf{R}_j$ are the atomic mass and coordinates of the $j$th atom in the supercell. This projection assumes that phonon quasi-particles at finite temperature share the same polarization vectors with the corresponding harmonic phonons, in accordance with the standard many-body theory that the $\hat{{e}}_{\lambda}$ can be chosen as an unperturbed basis and each phonon quasi-particle acquires a self-energy through anharmonic interactions \cite{Maradudin,Cowley}. The anharmonic effects are taken into account by the frequency shift $\Delta \omega_{\lambda}$ and the linewidth $\Gamma_{\lambda}$, corresponding to the real and imaginary parts of the phonon self-energy respectively.

For a well-defined phonon quasi-particle, the VAF of a normal mode oscillates and decays in a finite correlation time. Then the power spectrum of a phonon quasi-particle can be obtained directly by the Fourier transform of the VAF \cite{Lu2017,Lu2018},
\begin{equation}
G_{\lambda}(\omega)= \int_{0}^{\infty}\langle v^{*}(0)v(t)\rangle_{\lambda} \textup{exp}(i\omega t)dt.
\end{equation}
The typical power spectrum for a normal mode follows a Lorentzian-type line shape, from which both the quasi-particle frequency $\widetilde{\omega}_{\lambda}$ and the linewidth $\Gamma_{\lambda}$ can be extracted.

\subsection{Lattice thermal conductivity}
Phonon heat transport in anisotropic materials can be described by the Fourier$'$s equation,
\begin{eqnarray}
J=-\kappa_{ij}\nabla T,
\end{eqnarray}
where $J$ is the heat flux density, $\kappa_{ij}$ is the thermal conductivity tensor, and $\nabla T$ is the temperature gradient. In the framework of the quasi-particle gas model, phonons are treated as finite lifetime harmonics that interact only through instantaneous collisions. Under this approximation, each phonon normal mode can be regarded as a quasi-particle with energy $\hbar \omega_{\textbf{q},s}$ and velocity $v_{\textbf{q},s}$, thus we can  get the expression of
\begin{eqnarray}
J= \frac{1}{N_{q} \Omega} \sum_{\textbf{q},s} f \hbar \omega_{\textbf{q},s}v_{\textbf{q},s},
\end{eqnarray}
where $N_{q}$ is the number of $q$ points in the Brillouin zone, $\Omega$ is the volume of the unit cell and $f$ is the occupation number of phonons in an energy level (state), which deviates from the equilibrium Bose-Einstein distribution function $f_0$ due to the existence of a temperature gradient. To obtain the specific expression of $f$, we introduce the Boltzmann transport equation, which can describe the transport problems caused by phonons, electrons, or other particles quantitatively.

When the system reaches a dynamic equilibrium state driven by temperature, we can obtain the equation \cite{Peierls,Ziman}
\begin{eqnarray}
\frac{\partial f}{\partial t}\bigg|_{\textup{temp}} + \frac{\partial f}{\partial t}\bigg|_{\textup{scatt}}= 0.
\end{eqnarray}
The first term represents the deviation from the equilibrium state caused by temperature and the second term stands for the scattering of phonons.
Introducing the temperature gradient, Eq. (11) can be further written as
\begin{eqnarray}
\frac{\partial f}{\partial t}\bigg|_{\textup{scatt}} = -\frac{\partial f}{\partial t}\bigg|_{\textup{temp}} = -v_{\textbf{q},s}\nabla T \frac{\partial f}{\partial T}.
\end{eqnarray}
Since the deviation of phonon distribution from the equilibrium state $f_0$ is small, Eq. (12) can be solved by introducing a mean phonon lifetime $\tau_{\textbf{q},s}$ to linearize the scattering term
\begin{eqnarray}
\frac{\partial f}{\partial t}\bigg|_{\textup{scatt}} =  \frac{f- f_0}{\tau_{\textbf{q},s}}.
\end{eqnarray}
By combining Eqs. (12) and (13), we can obtain the expression of $f$ as
\begin{eqnarray}
f =  f_0 - v_{\textbf{q},s} \tau_{\textbf{q},s} \nabla T \frac{\partial f}{\partial T}.
\end{eqnarray}
Due to the small deviation introduced by the temperature gradient, the second term in Eq. (14) can be replaced by $- v_{\textbf{q},s} \tau_{\textbf{q},s} \nabla T \frac{\partial f_0}{\partial T}$.
By substituting Eq. (14) into Eq. (10), we can get
\begin{eqnarray}
J= \frac{1}{N_{q} \Omega} \sum_{\textbf{q},s} \hbar \omega_{\textbf{q},s}v_{\textbf{q},s}^{\alpha}(-v_{\textbf{q},s}^{\beta} \tau_{\textbf{q},s} \nabla T \frac{\partial f_0}{\partial T}).
\end{eqnarray}
By comparing Eqs. (9) and (15), we can obtain the expression of lattice thermal conductivity tensor $\kappa^{\alpha \beta}$ as
\begin{eqnarray}
\kappa^{\alpha \beta}=\sum_{\textbf{q},s} C_v(\textbf{q},s)  v^{\alpha}_{\textbf{q},s} v^{\beta}_{\textbf{q},s} \tau_{\textbf{q},s}
\end{eqnarray}
where $v^{\alpha(\beta)}_{\textbf{q},s}$ is the group velocity component along the $\alpha(\beta)$-direction; $\tau_{\textbf{q},s}$ is the phonon lifetime of mode ($\textbf{q},s$); $C_v(\textbf{q},s)$ is the mode specific capacity of the following expression,
\begin{eqnarray}
C_v(\textbf{q},s)=\frac{k_\textup{B}}{N_{q} \Omega} (\frac{\hbar \omega_{\textbf{q},s}}{k_B T})^{2} f_0 (f_0 + 1)
\end{eqnarray}
where $k_\textup{B}$ is the Boltzmann constant.

\subsection{DFT calculations}
Structure optimization and AIMD simulations were performed using density functional theory with the projected-augmented-wave method~\cite{Blochl1994}, as implemented in the VASP (version 6.3.0) package~\cite{Kresse1999}. The electron exchange and correlation potential was described using the generalized gradient approximation with the Perdew-Burke-Ernzerhof form \cite{Perdew1996}. A vacuum thickness of 20 \AA$~$was employed in the normal direction ($z$-axis) of monolayer ZrSe$_2$ to avoid the spurious interaction among periodic images. For structure optimization with the primitive cell, the 9$\times$9$\times$6 and 9$\times$9$\times$1 $k$-point meshes were used for the BZ integration~\cite{Monkhorst1976} for bulk and monolayer ZrSe$_2$ respectively. The plane wave cutoff energy was set to 350 eV. The harmonic force constant matrix was computed using the supercell and the finite displacement method with a displacement amplitude of 0.01 \AA. The 4$\times$4$\times$2 and 4$\times$4$\times$1 supercells were employed for bulk and monolayer ZrSe$_2$, respectively. The harmonic phonon dispersion relations and polarization vectors were then obtained using the Phonopy \cite{Togo2010} post-processing package. For all AIMD calculations, 4$\times$4$\times$2 and 4$\times$4$\times$1 supercells were used for bulk and monolayer ZrSe$_2$, respectively. The canonical ensemble (NVT) was utilized in all MD simulations, with temperature oscillations being controlled by the Nos\'e thermostat \cite{Nose}.

\section{III. Results and discussions}
\subsection{A. Training Procedure}
\begin{figure}
\includegraphics[width=0.9\columnwidth]{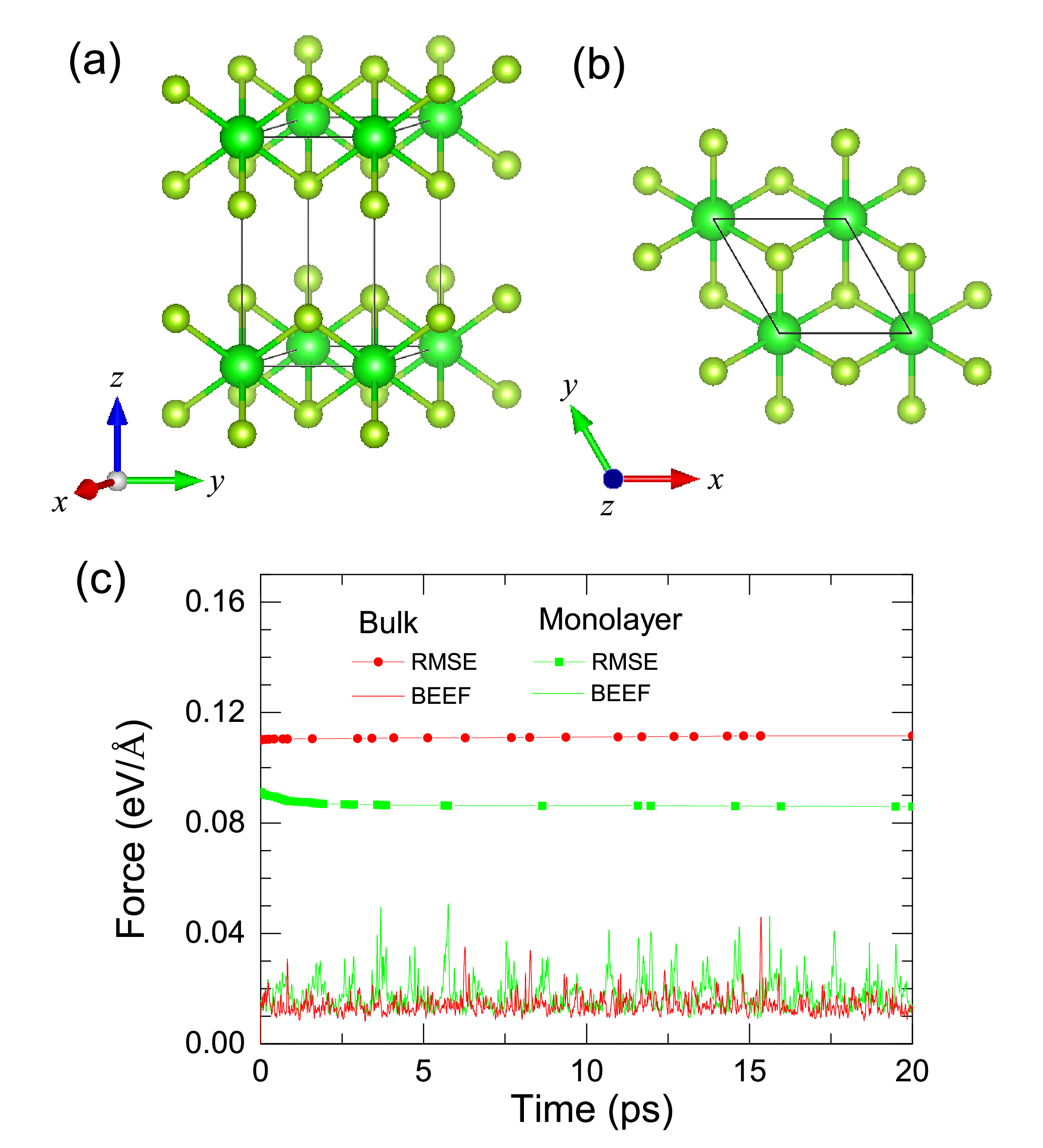}
\caption{(Color online) The crystal structure of bulk ZrSe$_2$ (a) in the side view and (b) in the top view. (c) The root-mean-square error of force (RMSE) and the Bayesian error estimation of force (BEEF) (eV/\AA) change over the last 20 ps of the force field prediction compared to the AIMD result.}
\end{figure}

The on-the-fly machine learning approach \cite{Jinnouchi2019,MLFF2} has been employed to enhance simulation speed. The VASP (version 6.3.0) package provides a comprehensive platform that combines force field potential training and FFMD simulations. Bayesian linear regression is utilized in the ridge regression method of machine learning to solve linear equations. The currently employed cutoff radius for the radial descriptor and angular descriptor in the MLFF method are 8.0 \AA$~$and 5.0 \AA, respectively. The width of the Gaussian functions used for broadening the atomic distributions in the radial descriptor and angular descriptor of the MLFF method is set at 0.5 \AA. To begin, we generated a 4$\times$4$\times$2 (96-atom) supercell for bulk ZrSe$_2$ and a 4$\times$4$\times$1 (48-atom) supercell for monolayer ZrSe$_2$. The training procedure was carried out in two stages. Initially, annealing was conducted using AIMD simulations, with a temperature range from 600 to 300 K over 80000 MD steps with a time step of 2 fs. In the second stage, the temperature range of 800 to 100 K, with temperature decrements of 100 K, was used. At each temperature, 20000 MD steps were performed using a time step of 2 fs, resulting in a total of 160000 AIMD steps (320 ps). Actually, in the training process, most of the AIMD steps were substituted with exceedingly rapid force-field steps.

The in-sample error, represented by the root-mean-square error (RMSE), denotes the average error within the training set. Conversely, the out-of-sample error or generalization error is the average error that arises when analyzing a new random configuration of the same ensemble. A well-trained force field should exhibit minimal out-of-sample error. The Bayesian error provides an evaluation of the out-of-sample error and tracks the progress of force field training. In Fig. 1(c), RMSE and Bayesian error estimation of force (BEEF) are displayed for the last 20 ps of force field predictions, as compared to AIMD results. Notably, both RMSE and BEEF demonstrate satisfactory convergence. For both bulk and monolayer phases, the RMSE of force converges to exceptionally small values of 0.11 and 0.09 eV/\AA, respectively. Correspondingly, the BEEF oscillates around 0.015 eV/\AA$~$for both phases, demonstrating excellent generalization error convergence. To assess the accuracy of the trained force field, we conducted a series of tests comparing it to AIMD results, including analysis of the pair correlation function $g$(r) and the probability distribution of atomic displacements. Fig. S1 in the Supplemental Material \cite{Supp} illustrates that the $g$(r) at 300 K calculated from the FFMD displays excellent agreement with AIMD results, with negligible errors. Fig. S2 and Fig. S3 in the Supplemental Material \cite{Supp} showcase the probability distribution of atomic displacements for Zr and Se atoms in the bulk and monolayer phases. At 300 K, all the Zr and Se atoms exhibit vibrations around the equilibrium lattice points, displaying a Gaussian function distribution. This highlights the stability of the bulk and monolayer phases at finite temperature \cite{Lu2019,Lu2021}. Furthermore, the distribution functions obtained from both FFMD and AIMD simulations demonstrated good agreement with each other, once again establishing the accuracy of the force field functions.

\subsection{B. Phonon quasiparticle}

\begin{figure}
\includegraphics[width=1.0\columnwidth]{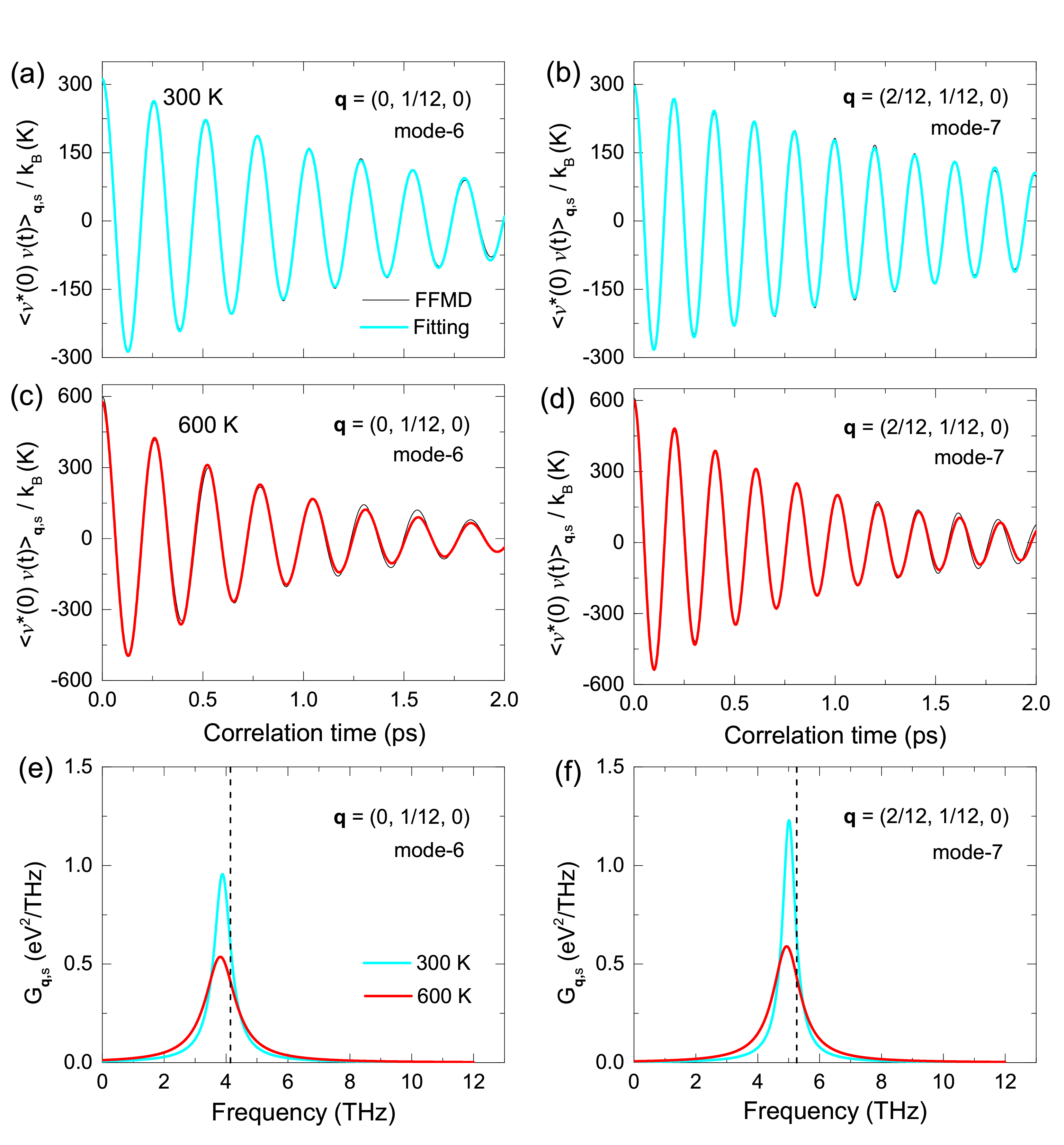}
\caption{(Color online) (a) The VAFs of mode-6 in ${\bf q}$=(0,$\frac{1}{12}$,0) at 300 K for monolayer ZrSe$_2$; (b) The VAFs of mode-7 at ${\bf q}$=($\frac{2}{12}$,$\frac{1}{12}$,0) at 300 K for monolayer ZrSe$_2$; (c) The same as (a) but for 600 K; (d) The same as (b) but for 600 K. The black curves are calculated directly by the mode-projected velocities based on the FFMD simulations. The cyan/red curves are obtained by fitting the VAFs according to Eq. (5). (e)-(f) The power spectrum, $G_{\textbf{q},s}$, of mode-6 and mode-7 at 300 and 600 K. The dashed black line indicates the harmonic frequency of 4.14 THz for mode-6 and 5.26 THz for mode-7 respectively.}
\end{figure}

Bulk ZrSe$_2$ crystalizes in the CdI$_2$ structure with the P$\bar{3}$m1 (164) space group, where the primitive cell contains one Zr and two Se atoms. The layers consist of a Se-Zr-Se sandwich-like atomic sequence, as presented in Fig. 1(a)-(b). Due to the weak van der Waals interactions between layers, the monolayer ZrSe$_2$ can be directly extracted from the bulk. The optimized lattice parameters are $a$=$b$=3.79 \AA$~$ and $c$=6.66 \AA$~$ for the bulk phase, and $a$=$b$=3.80 \AA$~$ for the monolayer ZrSe$_2$, consistent with previous experimental \cite{Whitehouse1978,Brauert1995} and theoretical \cite{Guo2014,Gao2021,Zhen2017} values. Phonon quasiparticles can be directly observed and defined using the VAF and the power spectrum obtained from the FFMD simulation. As such, we calculated the VAFs for all the normal modes using Eq. (6). Fig. 2 shows the VAFs of two representative normal modes, namely mode-6 at $\textbf{q}$=(0, $\frac{1}{12}$, 0) and mode-7 at $\textbf{q}$=($\frac{2}{12}$, $\frac{1}{12}$, 0), of monolayer ZrSe$_2$ at 300 and 600 K, respectively. All VAFs exhibit damping oscillations over time, indicating well-defined phonon quasi-particles. We extract phonon quasi-particle frequencies and linewidths by fitting these VAFs to an exponentially decaying cosine function following Eq. (5).

For mode-6 at $\textbf{q}$=(0, $\frac{1}{12}$, 0), the fitting curve with parameters ($\widetilde{\omega}_{\textbf{q},s}$, $\Gamma_{\textbf{q},s}$) of (3.88 THz, 0.67 THz) concurs well with the VAF produced by the FFMD simulation at 300 K, as presented in Fig. 2(a). Similarly, for mode-7 at $\textbf{q}$=($\frac{2}{12}$, $\frac{1}{12}$, 0), the corresponding fitting values are (5.01 THz, 0.52 THz). As temperature rises to 600 K, a significantly greater damping rate is observed for both modes, implying a shorter phonon quasiparticle lifetime. The fitted ($\widetilde{\omega}_{\textbf{q},s}$, $\Gamma_{\textbf{q},s}$) values of mode-6 and mode-7 are (3.81 THz, 1.19 THz) and (4.93 THz, 1.08 THz) respectively. Notably, the phonon power spectrum $G_{\textbf{q},s}$ in phase space that is directly obtained by the Fourier transform of the VAF using Eq. (5), can also extract the quasiparticle frequency and linewidth. Fig. 2(e)-(f) illustrate the power spectra of these two modes at 300 and 600 K, respectively. For a well-defined phonon quasiparticle, its power spectrum has a Lorentzian line shape \cite{Zhang2014,Lu2017}. As illustrated in Fig. 2(e)-(f), all power spectra exhibit a single-peak Lorentzian line shape, indicating the soundness of the phonon quasiparticle concept. The renormalized phonon frequency and linewidth correspond to the peak position and the full width at half maximum (FWHM), respectively. The extracted ($\widetilde{\omega}_{\textbf{q},s}$, $\Gamma_{\textbf{q},s}$) values of mode-6 are (3.88 THz, 0.67 THz) for 300 K and (3.81 THz, 1.18 THz) for 600 K, in good agreement with the values obtained by the fitting approach, which corroborates the validity and precision of the fitting approach. For mode-7, the frequency and linewidth obtained by both methods are equally consistent. Compared to the harmonic frequency of 4.14 and 5.26 THz, the frequency of both modes shows a red shift due to the anharmonic effect. With an increase in temperature, the FWHM of the curve increases, leading to a decrease in lifetime. As temperature increases from 300 to 600 K, the lifetime is reduced from 0.75 to 0.42 ps for mode-6 and from 0.96 to 0.46 ps for mode-7, respectively. Fig. S4 and Fig. S5 in the Supplemental Material \cite{Supp} also provide the fitting outcomes for other normal modes, namely all nine normal modes at $\textbf{q}$=($\frac{1}{4}$,0,0) in bulk and monolayer ZrSe$_2$. In this manner, the frequencies and lifetimes of all normal modes sampled in the BZ can be determined.

\subsection{C. Lattice thermal conductivity}
\begin{figure}
\includegraphics[width=0.8\columnwidth]{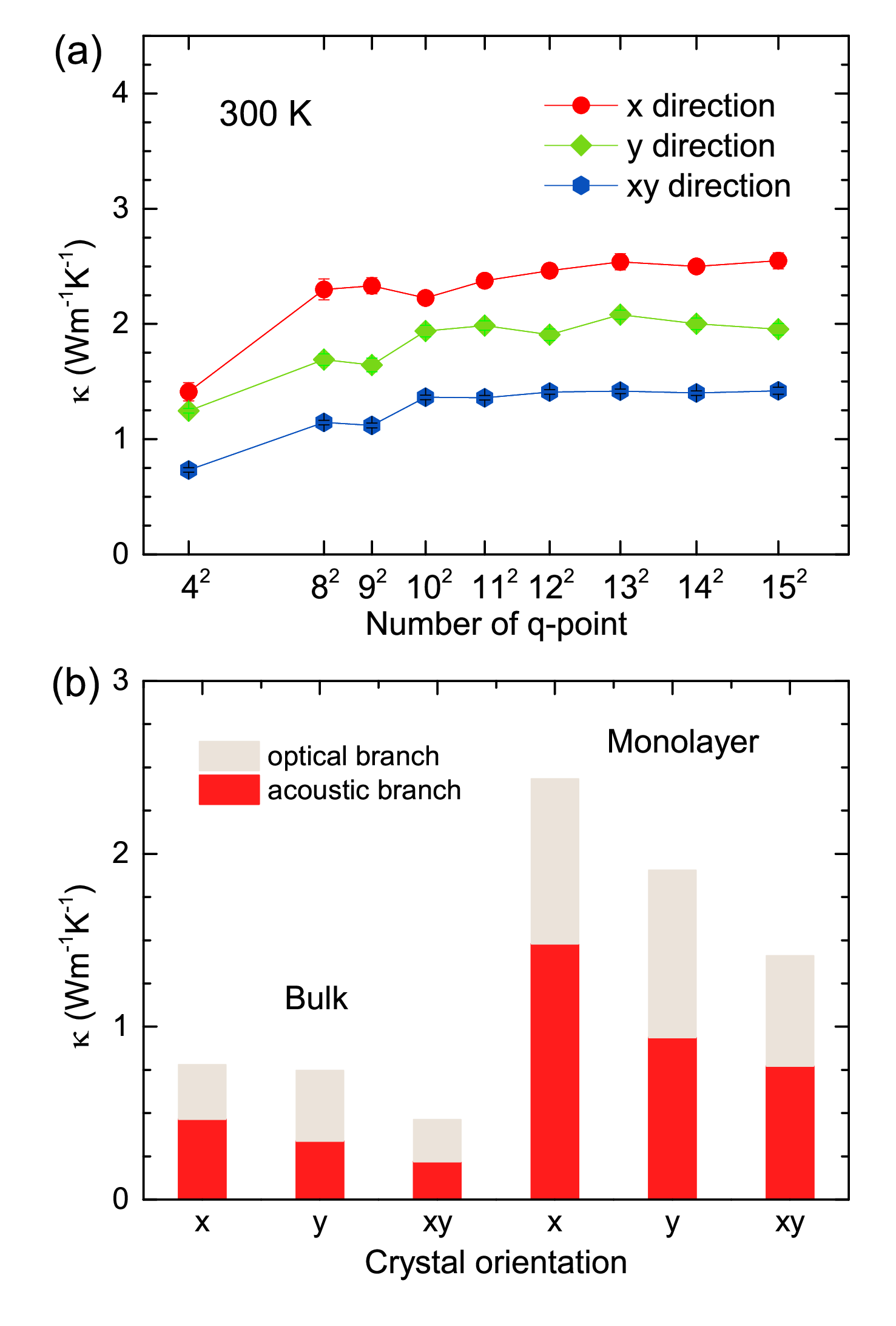}
\caption{(Color online) (a) The dependence of lattice thermal conductivity $\kappa$ (Wm$^{-1}$K$^{-1}$) on the number of $q$-points in the BZ of monolayer ZrSe$_2$ at 300 K. (b) The contribution of acoustic and optical phonon branches to the total lattice thermal conductivity using a $q$-point mesh of 12$\times$12$\times$12 for bulk phase and 12$\times$12$\times$1 for monolayer ZrSe$_2$ respectively.  }
\end{figure}

\begin{figure}
\includegraphics[width=0.8\columnwidth]{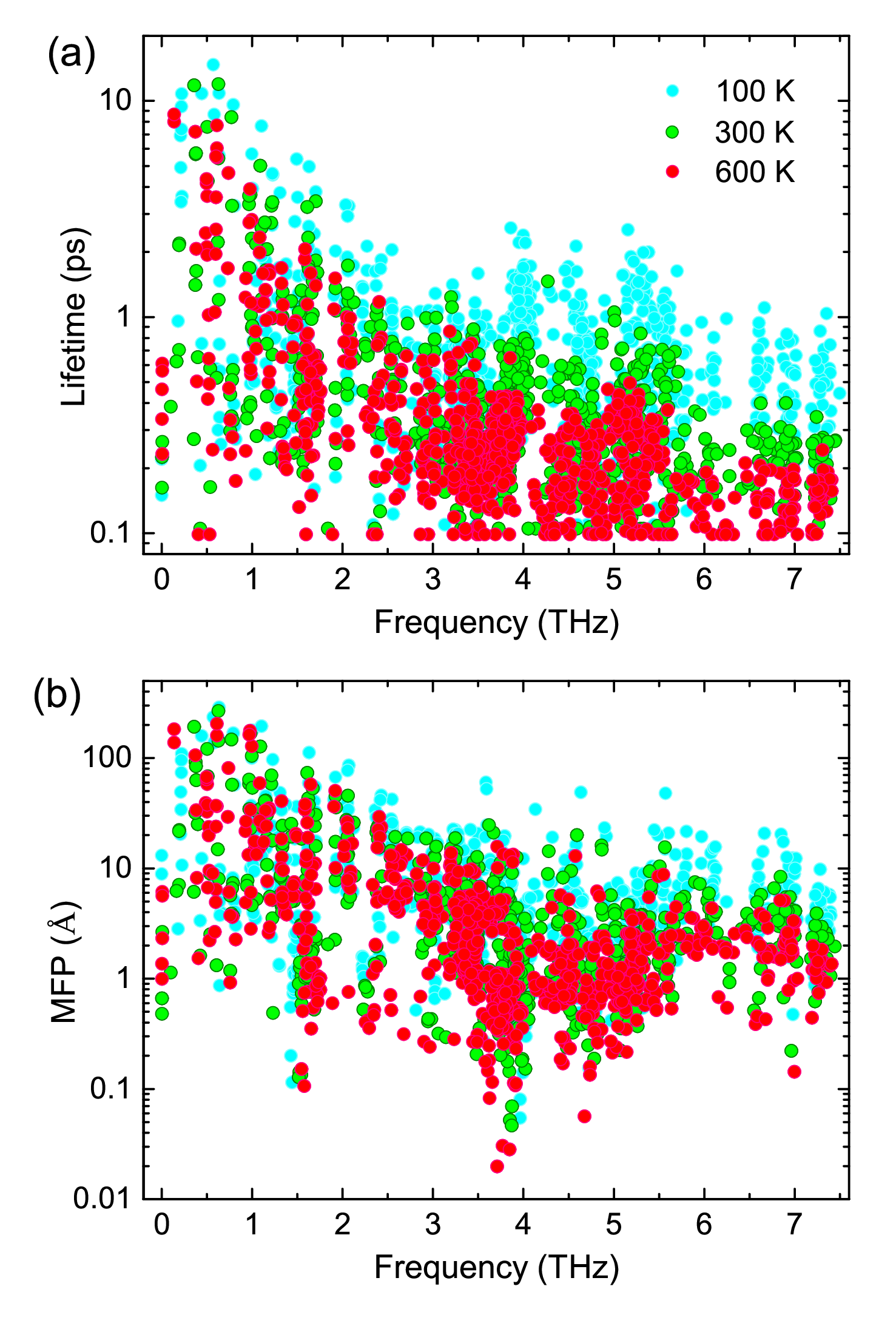}
\caption{(Color online) (a) The predicted phonon lifetime (ps) and (b) MFP (\AA) of monolayer ZrSe$_2$ as a function of frequency (THz) at 100  (cyan circles), 300 (green circles) and 600 K (red circles) respectively.}
\end{figure}

\begin{figure}
\includegraphics[width=0.9\columnwidth]{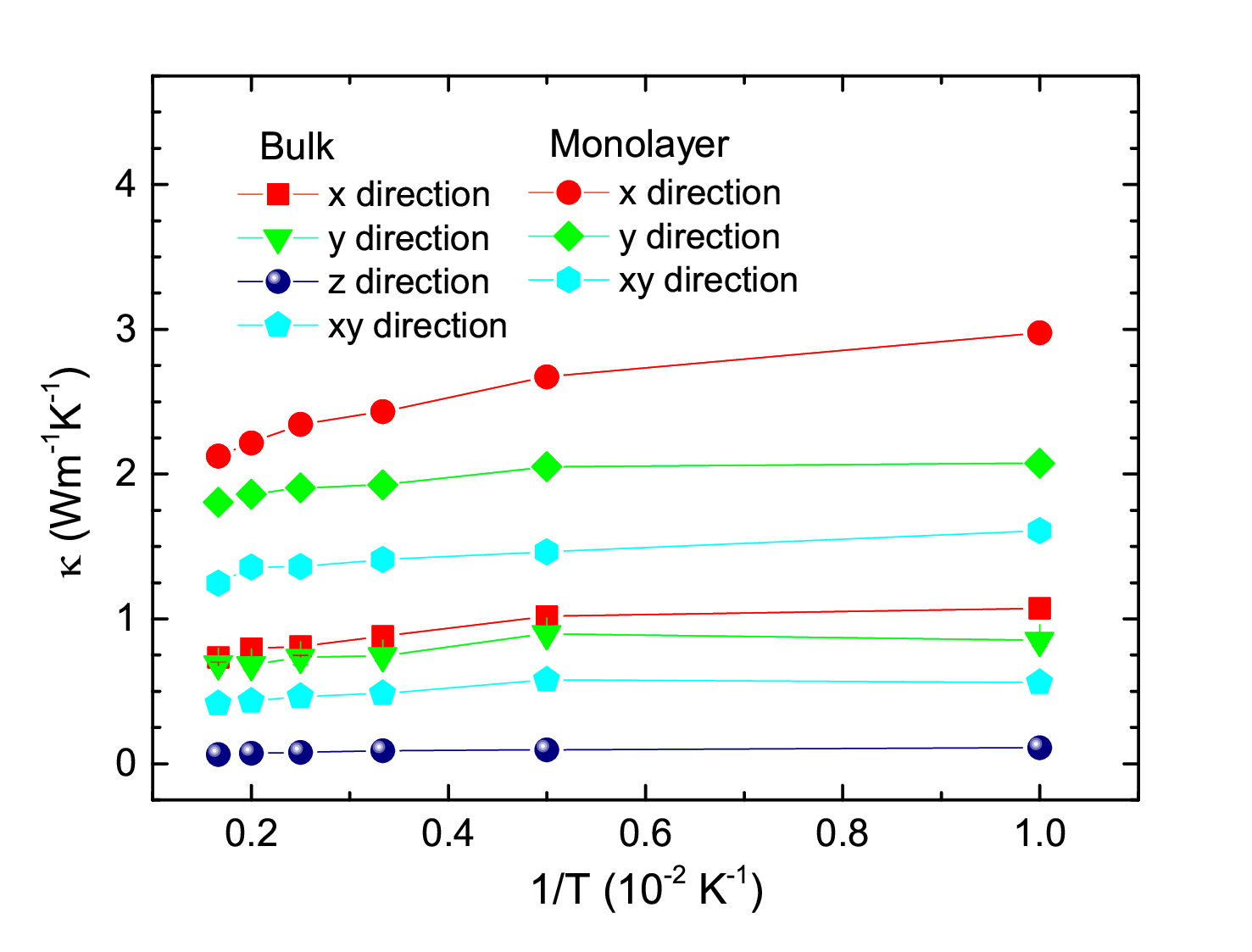}
\caption{(Color online) (a) The predicted lattice thermal conductivity tensor (Wm$^{-1}$K$^{-1}$) of bulk and monolayer ZrSe$_2$ as a function of inverse temperature $\frac{1}{T}$ (10$^{-2}$ K$^{-1}$). The temperature ranges from 100 to 600 K. With increasing temperature, $\kappa$ decays faster with temperature than the well-known $\frac{1}{T}$ scale.}
\end{figure}

\begin{figure}
\includegraphics[width=0.9\columnwidth]{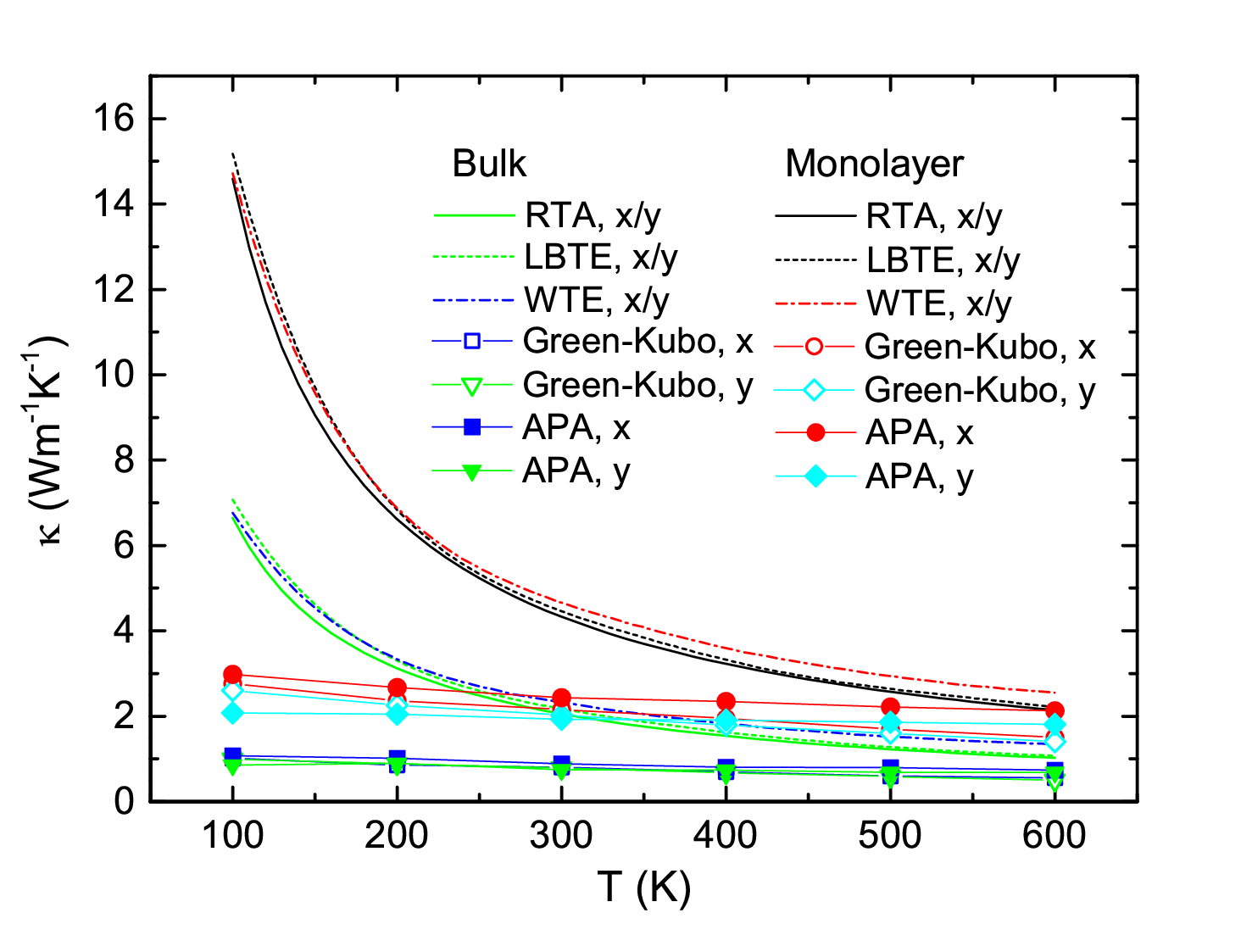}
\caption{(Color online) A comparison of the calculated LTC of bulk and monolayer ZrSe$_2$ using different methods: the relax-time approximation (RTA), the linearized phonon Boltzmann equation (LBTE), the Wigner transport equation (WTE), the Green-Kubo equation, and the present anharmonic phonon approach (APA).}
\end{figure}

The MLFF largely enhances the scale of simulations, which ensures sufficient $q$-points in the BZ to yield accurate results for LTC calculations. The convergence of LTC with respect to the supercell size or $q$-point mesh in the BZ has been checked. In MD simulations, the monolayer ZrSe$_2$ offers a lower dimensional advantage compared to the bulk phase. Additionally, it provides higher computational efficiency at the same scale. Therefore, we selected the monolayer ZrSe$_2$ to evaluate the convergence of supercells ($q$-point mesh). We constructed a supercell ranging from 4$\times$4$\times$1 to 15$\times$15$\times$1 for the monolayer ZrSe$_2$ to conduct the FFMD simulations. This corresponds to a $q$-point range of 4$^{2}$ to 15$^{2}$ in the BZ for calculating the LTC.

As shown in Fig. 3(a), the LTC curve increases as the number of $q$-points increases. For a $q$-point mesh of 4$^2$, which is typically used in AIMD \cite{Lu2021,Xue2022}, the calculated LTC values are 1.41, 1.25 and 0.73 Wm$^{-1}$K$^{-1}$ along the $x$, $y$ and $xy$ directions, respectively. It is important to note that the LTC value for this $q$-point mesh has not yet reach convergence. As the density of $q$-points increases, the LTC value significantly increases. All the predicted LTC values converge at around a 10$^{2}$ $q$-point mesh. Further increasing the supercell size, the change in LTC is minimal along the three crystal directions. In general, a $q$-point mesh of 12$^{2}$ can effectively balance accuracy and computational scale, resulting in LTC values of 2.46, 1.90 and 1.41 Wm$^{-1}$K$^{-1}$ along the $x$, $y$ and $xy$ directions, respectively. Increasing the $q$-point mesh to 15$^2$, the differences in LTC are less than 5 \%. Therefore, for subsequent calculations, we selected a 12$\times$12$\times$1 supercell (432 atoms) to calculate the LTC of the ZrSe$_2$ monolayer. Since the bulk and monolayer phases have identical atomic structure environments, we used the same scale as the monolayer for the bulk phase ZrSe$_2$, namely a 12$\times$12$\times$12 supercell (5184 atoms).

Compared with the pure MD approach for calculating LTC, the present method allows us to evaluate the contribution of each normal mode to LTC based on the BTE. By projecting the real-space VAF onto all individual modes with ($\textbf{q}$, $s$), we can effectively evaluate the contribution of each normal mode to LTC. In Fig. 3(b), it is shown that along the $x$ direction, the acoustic branch plays a major contribution to the LTC for both bulk and monolayer ZrSe$_2$. Specifically, in the bulk phase, the mode-3 (LA) branch contributes the most to the LTC with a contribution of 35.2\%. In the monolayer ZrSe$_2$, the mode-2 (TA$_2$) branch and mode-3 (LA) branch play a major role in the LTC with contributions of 25.3 \% and 21.8 \% respectively, as shown in Fig. S6 in the Supplemental Material \cite{Supp}. Along the $y$ direction, the contribution of optical branches to LTC is slightly higher than that of acoustic branches for bulk and monolayer ZrSe$_2$. The mode-5 (LO) branch becomes significant in its contribution, contributing 30.6 \% to the total LTC in the bulk phase. The LTC along the $xy$ direction is lower than that in the $x$ and $y$ directions for both phases, where acoustic and optical branches contribute almost equally to the LTC. The mode-6 (TO) mode contributes negligibly to LTC in both bulk and monolayer ZrSe$_2$, mainly due to the flat band property of its phonon dispersion relation as shown in Fig. S7 in the Supplemental Material \cite{Supp}, resulting in an extremely low phonon group velocity.

The phonon lifetime and MFP are temperature dependent in monolayer ZrSe$_2$. Fig. 4 illustrates the frequency-dependent phonon lifetime and MFP at temperatures of 100, 300 and 600 K. As the temperature increases, the phonon lifetimes decrease due to stronger anharmonic interaction, particularly for the optical modes. The cutoff frequency of the acoustic branch is approximately 3.5 THz (Fig. S7), consistent with the Debye temperature of 175 K for ZrSe$_2$. This temperature was calculated using density functional perturbation theory (DFPT) based on the stress-strain relationship of elastic strain \cite{Anderson1963,Voigt1928,Reuss1929,Hill1952,Baroni1987}. In general, phonon lifetimes decrease with increasing phonon frequencies following the relation $1/\tau_{\textbf{q},s} \propto \widetilde{\omega}_{\textbf{q},s}^{2}$. All phonon lifetimes are distributed below 20 ps. Optical modes are more sensitive to temperature, with lifetimes concentrated below 3 ps at 100 K and decreasing to within 1 ps at 600 K. This decrease in lifetimes is accompanied by a decrease in the MFP of phonons, as shown in Fig. 4(b). The distribution of phonon group velocity $v_g$ indicates that there is no significant difference in the $v_g$ between the acoustic branch and the optical branch (Supplemental Material Fig. S8). Therefore, the difference in phonon MFPs is primarily caused by the phonon lifetime. Most MFPs of optical modes are less than 40 \AA, which is within the simulation scale of the FFMD supercell, approximately 45.6 \AA$~$for both bulk and monolayer ZrSe$_2$.

According to Eq. (16), we have calculated the LTC tensor of both bulk and monolayer ZrSe$_2$ in the temperature range of  100 $<T<$ 600 K. The results of $\kappa$ as a function of $\frac{1}{T}$ (inverse temperature) are shown in Fig. 5. Notably, with the increasing of temperature, the LTC decays faster than the well-known $\frac{1}{T}$ scaling, which is typically observed when considering only three-phonon scattering. This behavior is due to the nature of the FFMD simulation, which effectively captures anharmonic effects up to infinite orders. Traditionally, the LTC between planes cannot be predicted using the BTE method based on the DFPT due to computational limitations. However, by taking advantage of the scalability offered by FFMD, our current approach can effectively predict the interlayer LTC of bulk ZrSe$_2$, as shown in Fig. 5. The interlayer LTC values, ranging from approximately 0.15 to 0.25 Wm$^{-1}$K$^{-1}$ between 100 and 600 K, are consistent with experimental results and general understanding of layered materials. This ultra-low LTC is attributed to the combination of extremely low phonon group velocity and phonon lifetime along the interlayer direction.

To provide a comprehensive comparison, we have subsequently employed several well-established approaches commonly used in the research community to calculate the LTC of bulk and monolayer ZrSe$_2$. These methods include the relaxation time approximation (RTA), the linearized phonon Boltzmann equation (LBTE) \cite{LBTE}, and the Wigner transport equation (WTE) \cite{WTE1,WTE2} based on three-phonon processes. Additionally, the LTC is also calculated via the Green-Kubo method \cite{Green1,Green2,Kubo1,Kubo2} based on the heat flux autocorrelation function from FFMD. To provide more clarity, we have included detailed calculation methodologies and intricacies in the Supplemental Material \cite{Supp}. In Fig. 6, all these results have been contrasted with our results. We can clearly observe the contribution of higher-order anharmonic phonon processes to LTC by comparing the results obtained considering only three-phonon processes with those including higher-order phonon processes. The LTC obtained by RTA, LBTE and WTE methods is significantly higher than that obtained using the Green-Kubo method and the present anharmonic phonon approach, especially below the Debye temperature of 175 K for ZrSe$_2$. Our results agree well with those obtained from the Green-Kubo method, both of which involve higher-order phonon processes introduced by the FFMD. By studying the heat flux autocorrelation function (Supplemental Material Fig. S9 \cite{Supp}), we observe a clear increase in the thermal dissipation rate with temperature. The consideration of higher-order phonon processes significantly enhances the phonon scattering rate, leading to a reduction in LTC. Additionally, we demonstrate the effect of phonon anharmonicity on vibrational entropy \cite{Dove1993}. As shown in the Supplemental Material Fig. S10 \cite{Supp}, our results reveal a significant increase in the difference between the vibrational entropies obtained using the anharmonic phonon approach and the harmonic approximation method as the temperature increases. These findings highlight the importance of higher-order anharmonic phonon processes in governing thermal transport.

\section{IV. Summary}
In this paper, the LTC of layered transition metal disulfide ZrSe$_2$ is theoretically calculated in combination with machine learning FFMD calculations and the Boltzmann transport equation. The use of FFMD effectively expanded the computational scale of AIMD calculations and broadened the range of calculation methods available for lattice thermal conductivity. Molecular dynamics has methodological advantages in considering phonon anharmonic effects, encompassing all order anharmonic interactions between phonons, and thereby effectively avoiding the explosion of calculation dimension caused by high-order force constant terms in the potential function series expansion method. At the same time, this method can quantify the anharmonicity of each normal mode, which is of great value in analyzing physical phenomena such as high-temperature phase transitions and superconductivity caused by specific normal modes. The phonon lifetime and mean free path of ZrSe$_2$ exhibit significant dependence on temperature.
The decrease in the LTC tensor with increasing temperature surpasses the typical $\frac{1}{T}$ scale, due to the inclusion of higher-order phonon anharmonic interactions, which is consistent with the results of the Green-Kubo method. The contribution of each individual normal mode to the LTC of ZrSe$_2$ is successfully traced based on the BTE, which is helpful for accurate control of the LTC. This method has universality in calculating LTC in bulk and low-dimensional crystals, as well as interlayer LTC of layered materials.

This work is supported by the National Natural Science Foundation of China under Grant Nos. 12074028, 12022415 and 11974056.

\end{document}